\documentclass[fleqn,twoside]{article}
\usepackage[headings]{espcrc2}
\usepackage{epsfig}

\readRCS
$Id: Rubbia_nuint04.tex,v 1.1 2004/09/08 13:30:40 rubbiaa Exp $
\usepackage{graphicx}
\usepackage[figuresright]{rotating}

\newcommand{\AmS}{{\protect\the\textfont2
  A\kern-.1667em\lower.5ex\hbox{M}\kern-.125emS}}

\title{Ideas for future liquid Argon detectors\thanks{Invited talk at the Third International Workshop on 
Neutrino-Nucleus Interactions 
in the Few GeV Region, NUINT04, March 2004, Gran Sasso, Italy}}

\author{A. Ereditato\address[MCSD]{INFN Napoli, Naples, Italy}
                and
        A. Rubbia\address[ETHZ]{Institut f\"{u}r Teilchenphysik, ETHZ, CH-8093 Z\"{u}rich,
Switzerland}}
       
\runtitle{2-column format camera-ready paper in \LaTeX}
\runauthor{S. Pepping}

\begin{document}

\begin{abstract}
We outline a strategy for future 
experiments on neutrino and astroparticle physics 
based on the use, at different detector mass scales (100 ton and 100 kton), of the liquid Argon Time Projection Chamber (LAr TPC) 
technique. The LAr TPC technology has great potentials for both cases
with large degree of interplay between the two applications and a strong synergy.
The ICARUS R\&D programme has demonstrated that the
technology is mature and that one can built a large ($\sim$ 1 kton) LAr TPC.
We believe that one can conceive and design a very large mass LAr TPC
with a mass of 100 kton by employing a monolithic technology based on the use of industrial,
large volume cryogenic tankers developed by the petro-chemical industry.
We show a potential implementation of a large LAr TPC detector. Such a detector
would be an ideal match for a Superbeam, Betabeam or Neutrino Factory, 
covering a broad
physics program that could include the detection of atmospheric, 
solar and supernova neutrinos, and search for
proton decays, in addition to the rich accelerator neutrino physics program.
In parallel, physics is calling for another application of the LAr TPC technique
at the level of 100 ton mass, for
low energy neutrino physics and for use as a near station
setup in future long baseline neutrino facilities.
We present here the main physics objectives and outline the conceptual design of
such a detector.\vspace{1pc}
\end{abstract}

\maketitle


\begin{figure*}[htb]
\centering
\epsfig{file=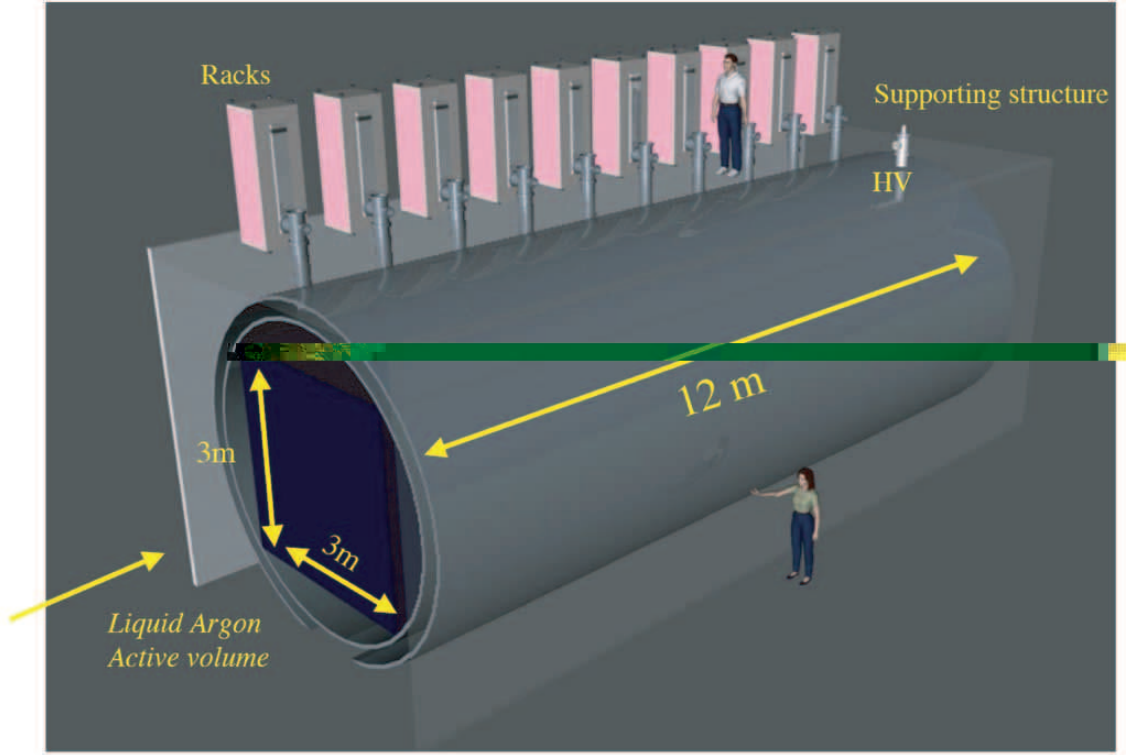,width=15cm}
\vspace*{-1cm}
\caption{\small Conceptual design of a 100~ton liquid Argon TPC detector.}
\label{fig:t150concept}
\end{figure*}

\section{The liquid Argon TPC technique}
Among the many ideas developed around the use of liquid noble gases, the Liquid 
Argon Time Projection Chamber (LAr TPC), conceived and
proposed at CERN by C.~Rubbia in 1977~\cite{intro1}, certainly represented one of the most
challenging and appealing designs.
The technology was proposed as a tool for
uniform and high accuracy imaging of massive detector volumes. 
The operating principle of the LAr TPC was based on
the fact that in highly purified LAr ionization tracks could indeed be transported
undistorted by a uniform electric field over distances of the 
order of meters. Imaging is
provided by wire planes placed at the end of
the drift path, continuously sensing and recording the signals induced by
the drifting electrons. Liquid Argon is an ideal medium since it provides
high density, excellent properties (ionization, scintillation yields) and
is intrinsically safe and cheap, and readily available anywhere as a standard by-product
of the liquefaction of air.

Non--destructive readout of ionization electrons by
charge induction allows to detect the signal of electrons crossing
subsequent wire planes with different orientation. This provides several
projective views of the same event, hence allowing for space point 
reconstruction and precise calorimetric measurement.

The detector performance can be summarized as
(1) a tracking device with precise event topology reconstruction;
(2) momentum estimation via multiple scattering;
(3) measurement of local energy deposition ($dE/dx$), providing
$e/\pi^0$ separation (sampling typ. $2\%X_0$), particle identification
via range versus $dE/dx$ measurement;
(4) total energy reconstruction of the event from charge integration 
(the volume can be considered as a full--sampling, fully homogenous calorimeter)
providing excellent accuracy for contained events.

The main technological challenges of this detection technique have been recently summarized
elsewhere~\cite{t600paper}. They mainly consisted in: (1) techniques of Argon purification, (2) operation of wire 
chambers in cryogenic liquid
and without charge amplification, (3) extremely low--noise analog electronics, (4) continuous wave--form recording
and digital signal processing.

The feasibility of this technology has been
demonstrated by the extensive ICARUS R\&D program, which included
studies on small LAr volumes about proof of principle, LAr purification
methods, readout schemes and electronics, as well as studies with
several prototypes of increasing mass on purification technology,
collection of physics events, pattern recognition, long duration tests and
readout. The largest of these devices had a mass of 3 tons of
LAr~\cite{3tons,Cennini:ha} and has been continuously operated for more than four years, collecting
a large sample of cosmic-ray and gamma-source events. Furthermore, a smaller
device with 50 l of LAr~\cite{50lt} was exposed to the CERN neutrino
beam, demonstrating the high recognition capability of the technique for
neutrino interaction events.

The realization of the 600 ton ICARUS detector culminated with its full test 
carried out at surface during the summer 2001~\cite{t600paper}. This test demonstrated that
the LAr TPC technique can be operated at the kton scale with a drift length of 1.5~m.
Data taking of about 30000 cosmic-ray triggers have allowed to test
the detector performance in a quantitative way and results have been 
published in~\cite{Amoruso:2003sw,Amoruso:2004dy,gg2,gg3,gg1}. Transportation
and installation at the Gran Sasso Underground Laboratory is currently
on-going.

\begin{figure*}[htb]
\centering
\epsfig{file=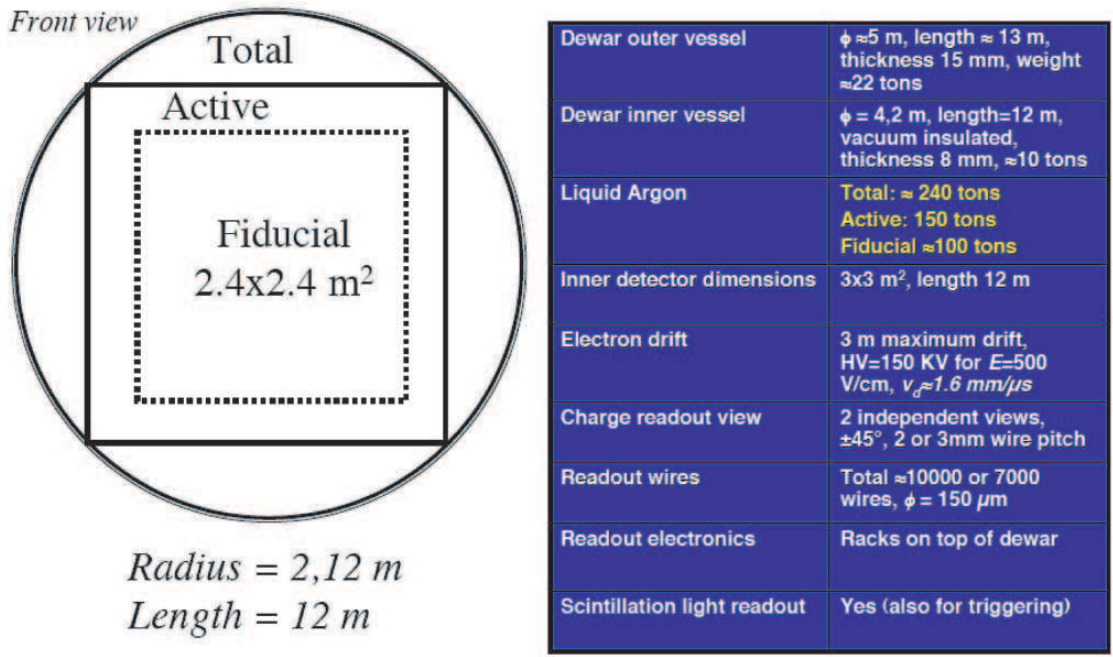,width=15cm}
\vspace*{-1cm}
\caption{\small Preliminary parameters of a 100~ton liquid Argon TPC detector.}
\label{fig:t150parameters}
\end{figure*}

\section {What the liquid Argon TPC provides}
As already mentioned, the bubble-chamber-like event reconstruction capability provide simultaneously (1) a tracking
device with unbiased imaging and reconstruction, and (2) a full sampling calorimetry.
The detector is fully active, homogeneous and isotropic. 
The energy resolution is very good, both for energy (calorimetry) and
for angular reconstruction (tracking).
The energy resolutions of contained particles are
$\sigma/E = 11\%/\sqrt{E(MeV)}\oplus 2\%$ for low energy electrons (measured~\cite{Amoruso:2003sw}),
$\sigma/E \approx 3\%/\sqrt{E(GeV)}$ for electromagnetic showers,
$\sigma/E \approx 30\%/\sqrt{E(GeV)}$ for hadronic  showers (pure LAr),
$\sigma/E \approx 17\%/\sqrt{E(GeV)}$ for hadronic showers (TMG doped LAr).
The chamber can be readily used in a broad energy range, from MeV up to multi-GeV's with
high event reconstruction efficiency. At the same time, low thresholds for particle identification
are possible given the high granularity. Combining $dE/dx$ measurements
and range it is possible to separate muons, pions, kaons, and protons.
Separation of electrons from neutral pions, and of muons from pions are
also highly efficient, owing to the imaging and multiple $dE/dx$ measurements.
Like bubble chambers, it is possible to operate a liquid argon TPC in 
a shallow depth (even though these are slow devices) owing to the high
granularity which permits the separation of the signal from the background
activity. Finally, it is possible to embed the detector in a magnetic
field for charge discrimination\cite{Rubbia:2001pk}.

Implementations at different mass scales (e.g. from 100~tons to 100~ktons)
are conceivable.

\begin{figure*}[htb]
\centering
\epsfig{file=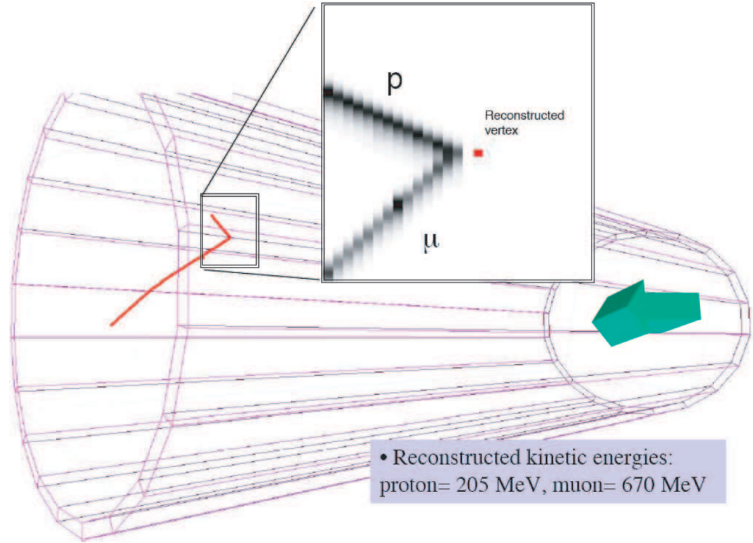,width=10cm}
\vspace*{-1cm}
\caption{\small Reconstruction of a quasi-elastic event in a 150~ton liquid Argon volume.}
\label{fig:t150qel}
\end{figure*}

\section {A 100 ton detector in a near station of a LBL experiment}

\subsection{Tentative design}
A 100 ton detector in a near-site of a long-baseline facility is a straight forward and
very desirable application of the technique. 
For example, the approved T2K experiment in Japan\cite{Itow:2001ee}
or other beams in the US might provide the ideal conditions to study with high statistical
accuracy neutrino interactions on liquid Argon in the very important energy range around 1 GeV.
This is a mandatory step in order to be able to handle high statistics provided by large detectors.
A prototype of a 100 ton detector could provide a tool to study calorimetric (electromagnetic
and hadronic) response in a charged particle beam or be readily placed in an existing
neutrino beam.

A tentative layout of the detector is shown in Figure~\ref{fig:t150concept}.
The active liquid Argon volume is envisaged to be $3\times 3\times 12$~m$^3$.
The liquid Argon is stored in a cylindrical cryostat with vacuum insulation.
Around the cryostat a supporting structure allows to install the electronic racks
close to the signal feed-throughs located at the top of the dewar. Other services
(e.g. high-voltage) are also located at the top.
A preliminary list of parameters is given in Figure~\ref{fig:t150parameters}.
The front view of the detector with the definition of the various liquid Argon
subvolumes is also shown.
A typical quasi-elastic muon neutrino interaction is shown in Figure~\ref{fig:t150qel}.

We discuss in the following sections in more details the case of the T2K experiment.

\subsection{The T2K project}
The JHF-Kamioka neutrino project is a second generation long baseline 
neutrino oscillation experiment that probes physics beyond the Standard
Model by high precision measurements of the neutrino masses and mixing.
A high intensity narrow band neutrino beam is produced 
by secondary pions created by a high intensity
proton synchrotron at JHF (JAERI).
The neutrino energy is tuned to the oscillation maximum 
at $\sim$1~GeV for a baseline length of 295~km towards the
world largest water \v{C}erenkov detector, Super-Kamiokande.

The project is divided into two phases. 
In the first phase, the main physics goal 
is the precision measurement of neutrino oscillation with the 50~GeV~PS of 
0.77~MW beam power and Super-Kamiokande. 
The physics goal of the first phase is 
an order of magnitude better precision
in the $\nu_\mu\rightarrow \nu_\tau$ oscillation measurement
($\delta(\Delta m_{23}^2)=10^{-4}$~eV$^2$ 
and $\sin^22\theta_{23}$ with $1 \%$ precision), 
a factor of 20 more sensitive search in the $\nu_\mu\rightarrow\nu_e$
appearance ($\sin^22\theta_{13}>0.006$),
and a confirmation of the $\nu_\mu\rightarrow \nu_\tau$ oscillation
or discovery of sterile neutrinos by detecting the neutral current
events. During a second phase, the power of the neutrino
beam could be increased and a new far detector could
be considered\cite{Itow:2001ee}.

\subsection{The measurements at the near stations of T2K}
The T2K long-baseline program foresees (1) one near station at 280~m, (2) an option for a second
intermediate station at 2~km and (3) the far station composed by the existing Superkamiokande detector.

In order to achieve the challenging goals of the T2K program, the following items will need to be addressed:
\begin{enumerate}
\item For the disappearance experiment: in order to determine precisely the $\Delta m_{23}^2$ and $\sin^22\theta_{23}$
parameters with small systematic errors, a very good knowledge of the neutrino beam will have to be reached.
\item For the appearance experiment: in order to understand precisely the beam associated backgrounds to the
electron appearance search, a very good knowledge of (a) the intrinsic $\nu_e$ component of the beam and
(b) the $\pi^0$ production in neutrino interactions in the GeV range will be mandatory.
\end{enumerate}

We assume that the near detectors will be composed of different technologies, like in the
case of the presently running K2K experiment\cite{Ahn:2002up}, including both
Water Cerenkov and ``fine grained'' detectors.
There are plans for a 1~kton Water Cerenkov detector and other fine grain detectors.
At the 2~km position, the rate in a 100~ton detector would be about 300'000 events per
year. This is a unique location for a liquid Argon TPC, since this technique would
allow to readily reach the required mass while keeping the fine granularity.

We give in the following a list of physics measurements that could be performed in the
near stations with a liquid Argon detector:

\noindent (A) Measurement of $\nu_\mu$~CC events: the liquid Argon TPC will provide an
independent measurement of the ``off-axis'' flux. The excellent muon identification
makes the selected sample very clean and the reconstruction will be unbiased.
The low momentum detection threshold in LAr compared to Water Cerenkov
allows for an independent classification and measurement of event samples
in the GeV region. This will provide independent systematic on the $nQE/QE$ ratio
and on the energy scale\cite{Ahn:2002up}. In addition, the independent reconstruction effects
in Water Cerenkov can be studied with the events recorded in the LAr and this
will help in the understanding the systematic errors in extrapolating the 1~kton
Water Cerenkov to SuperK. Finally, the energy independent detection and measurement
efficiency for subGeV and multiGeV events in LAr will provide a measurement of
the high energy muon neutrinos from kaon decays for an extra handle on the $\nu_e$
component of the beam.

\noindent (B)  Measurement of $\nu$~NC events: the clean measurement of the $\pi^0$ production
will provide an independent systematic error on the $NC/CC$ ratio. One can also
address independently the coherent production by looking for the absence of tracks
at the vertex. The clean $e/\pi^0$ separation available thanks to the excellent
event and particle identification plays here an important role.

\noindent (C)  Measurement of intrinsic $\nu_e$~CC events:  the excellent event and particle
identification give clean $e/\mu$ and $e/\pi^0$ separation with an unbiased
reconstruction. This will provide an independent measurement of the $\nu_e$ contamination,
well separated from the $\pi^0$ background. Combined with the NC background,
it will yield independent and separated components $\nu_e$ and $\pi^0$ background
at the far detector.

\noindent (D)  Standard model neutrino interactions in the GeV region: the bubble-chamber
like imaging will permit the study of neutrino interactions with high quality and given
the flux and large mass with high statistics. This sample of events will allow the study
of the DIS+resonance modeling, the QE modeling (form factors, ...), and
the nuclear effects (binding, Fermi-motion,
Pauli-exclusion, NN-correlations, PDF modifications, rescattering, ...).

We recall that the neutrino beam at JAERI is constructed as an ``off-axis'' beam with an angle between
2 and 3 degrees. While a lot of experimental knowledge has been acquired with the prediction of on-axis fluxes,
there is no previous experience in the operation of off-axis beam. In particular, we note that the systematics
associated to on- and off-axis beam could be quite different: (1) in an on-axis beam, the prediction of the neutrino
flux requires a good knowledge of the meson spectrum yield and to lesser extent their angular distribution since
the beam at the far location is very wide and the detector is placed at the maximum of the flux. On the other hand,
an off-axis beam requires the knowledge of the meson spectrum with less precision, since in the ideal case of the Jacobian
peak, the neutrino flux is actually independent of the parent meson momentum. Hence, the meson spectrum is less
important. On the other hand, angular effects are most important, as can be appreciated by considering the modification
in the neutrino spectrum when the beam angle is changed by 1 degree.

It appears natural that a conservative approach to the first ever performed off-axis experiment with the goals
of measuring very precisely the oscillation parameters will require some level of redundancy.

\begin{figure*}[htb]
\centerline{\epsfxsize=\textwidth\epsfbox{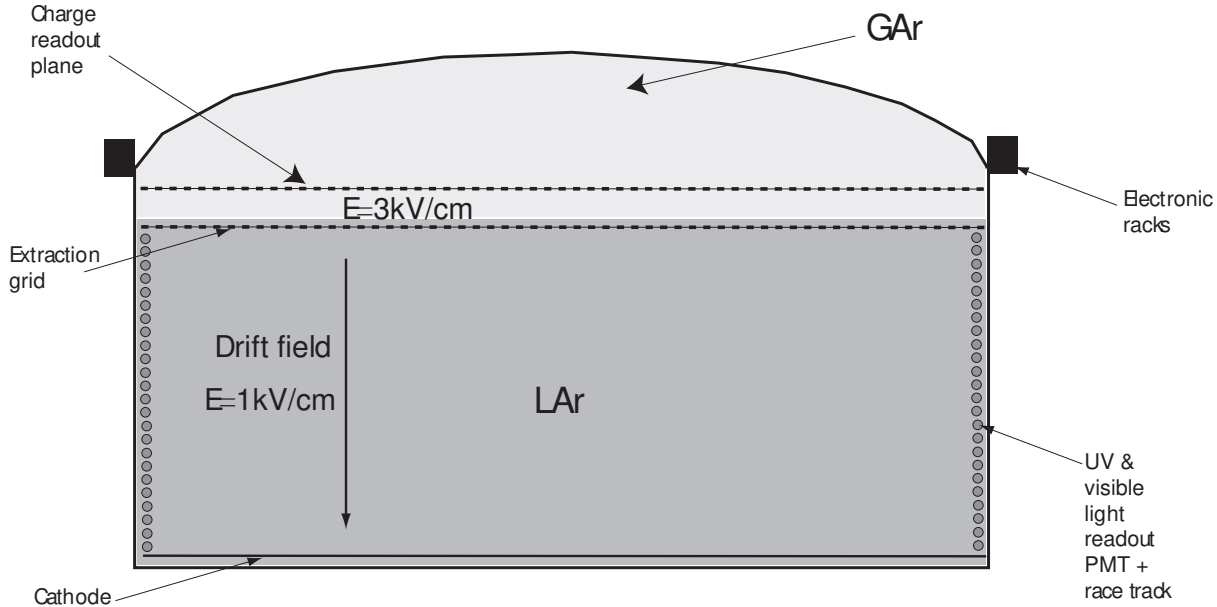}}   
\vspace*{-1cm}
\caption{Schematic layout of a 100 kton liquid Argon detector. The race track is composed of a set
of field shaping electrodes.}
\label{fig:t100schema}
\end{figure*}

\section {A 100 kton liquid Argon TPC detector with charge imaging, scintillation and Cerenkov light readout}

A 100 kton liquid Argon TPC would deliver extraordinary physics output 
(sometimes called ``megaton physics"), owing to better event reconstruction
capabilities provided by the LAr technique. A 100~kton
LAr TPC would represent one of the most advanced massive underground detectors
built so far~\cite{Rubbia:2004yq}. 

\subsection{Tentative design}
A conceptual design for a 100~kton LAr TPC was first given in Ref.~\cite{Rubbia:2004tz}.
The basic design features of the detector can be summarized as follows:
(1) Single 100 kton ``boiling'' cryogenic tanker at atmospheric
pressure for a stable and safe equilibrium condition (temperature is constant while Argon is boiling).
The evaporation rate is small (less than $10^{-3}$ of the total volume per day given
by the very favorable area to volume ratio) and is compensated
by corresponding refilling of the evaporated Argon volume. 
(2) { Charge imaging, scintillation and Cerenkov light readout}
for a complete (redundant) event reconstruction. This represents a clear advantage over large
mass, alternative detectors operating with only one of these readout modes. The physics
benefit of the complementary charge, scintillation and Cerenkov readout
are being assessed.
(3)  { Charge amplification to allow for very long drift paths}. 
The detector is running in bi-phase mode. In order to allow for drift lengths as long as $\sim$ 20 m,
which provides an economical way to increase the volume of the detector with a constant number
of channels, charge attenuation will occur along the drift due to attachment to the remnant impurities present
in the LAr. This effect can be compensated with 
charge amplification near the anodes located in the gas phase.
(4)  { Absence of magnetic field}, although this possibility might be considered at a later
stage. Physics studies~\cite{Rubbia:2001pk} indicate that a magnetic field
is really only necessary when the detector is coupled to a Neutrino Factory.

The cryogenic features of the proposed design are based on the industrial
know-how in the storage of liquefied natural gases (LNG, $T\simeq 110$ K at 1 bar),
which developed quite dramatically in the last decades, driven by the petrochemical and space rocket industries. 
LNG are used when volume is an issue, in particular, for storage.
The technical problems associated to the design of large cryogenic tankers,
their construction and safe operation have already been addressed and engineering problems
have been solved by the petrochemical industry. 
The current state-of-the-art contemplates cryogenic tankers of
200000~m$^3$ and their number in the world is estimated to be $\sim$~2000
with volumes larger than 30000~m$^3$ with the vast majority built
during the last 40 years. 
Technodyne International Limited, UK~\cite{Technodyne}, which
has expertise in the design of LNG tankers, has been appointed to initiate a feasibility
study in order to understand and clarify the issues related to the operation of a large
underground LAr detector. A final report is expected soon.

A schematic layout of the inner detector is shown in Figure~\ref{fig:t100schema}. The detector is characterized
by the large fiducial volume of LAr included in a large tanker, with external dimensions
of approximately 40 m in height and 70 m in diameter. A cathode located at the bottom of the 
inner tanker volume 
creates a drift electric field of the order of 1~kV/cm over a distance of about 20~m. 
In this field configuration ionization electrons
are moving upwards while ions are going downward. The electric field is delimited on the sides of the tanker
by a series of ring electrodes (race-tracks) placed at the appropriate potential by a voltage divider.

The tanker contains both liquid and gas Argon phases at equilibrium. Since purity is a concern for very long
drifts of 20 m, we assume that the inner detector could be operated in bi-phase mode:
drift electrons produced in the liquid phase are extracted from the liquid into the gas phase with
the help of a suitable electric field and then amplified near the anodes. 
In order to amplify the extracted charge one can consider various options: amplification
near thin readout wires, GEM~\cite{Sauli:qp} or LEM~\cite{Jeanneret:mr}. 
Studies that we are presently conducting show that gain factors of 100-1000 are achievable in pure Argon~\cite{dmrd}.
Amplification operates in proportional mode. Since the readout is limited to the top of the detector, 
it is practical to route cables out from
the top of the dewar where electronics crates can be located around the dewar outer edges.

After a drift of 20 m at 1 kV/cm, the electron cloud
diffusion reaches approximately a size of 3 mm, which corresponds to the
envisaged readout pitch. Therefore, 20 m practically
corresponds to the longest conceivable drift path. 
As mentioned above, drifting over such distances
will be possible allowing for some charge attenuation due to attachment
to impurities. If one assumes that the operating electron lifetime is at least $\tau\simeq 2$~ms (this is
the value obtained in ICARUS~T600 detector during the technical run~\cite{gg3} and
better values of up to $10$~ms were reached on smaller prototypes during longer runs),
one then expects an attenuation of a factor $\sim$ 150 over the distance of 20~m. 
This loss will be compensated by the proportional gain at the anodes.
We remind that the expected attenuation factor (compensated by the amplification) will not introduce
any detection inefficiency, given the value of $\sim$ 6000 ionization electrons per millimeter produced 
along a minimum ionizing track in LAr.

In addition to charge readout, one can envision to locate PMTs around the inner surface of the tanker. 
Scintillation and Cerenkov light can be readout essentially independently. 
LAr is a very good scintillator with about 50000 $\gamma$/MeV (at zero electric field). 
However, this light is essentially distributed around a line at $\lambda=128$~nm and, therefore, 
a PMT wavelength shifter (WLS) coating is required. 
Cerenkov light from penetrating muon tracks has been successfully detected in a LAr TPC~\cite{gg2};
this much weaker radiation (about $700\ \gamma/$MeV between 
160~nm and 600~nm for an ultrarelativistic muon) can be separately identified with
PMTs without WLS coating, since their efficiency for the DUV light
will be very small. 

\subsection{R\&D towards a 100~kton liquid Argon detector}

Studies are underway and planned
with the aim of identifiying the main issues 
of the future systematic R\&D and optimization activities\cite{multimw}: 

\noindent (1) {\bf The study of suitable charge extraction, amplification and collection devices}:
We are continuing an R\&D study to further optimize the technique for charge extraction, amplification and collection.
We are seeking a solution which yields gains between 100 and 1000 in pure Argon, which is electrically
and mechanically stable, and easy to be mass produced.

\noindent (2) {\bf The understanding of charge collection under high pressure as expected for events occurring at the bottom of the
large cryogenic tanker}:
We are constructing a small chamber which will be pressurized to 3-4~bar to simulate
the hydrostatic pressure present at the bottom of a future 100~kton tanker. We intend
to check that the drift properties of electrons are not affected at these pressures.

\noindent (3) {\bf The realization of a 5 m long detector column}:
We are constructing a column-like dewar 6 m long and 40 cm in diameter which will contain a 5 m long prototype
LAr detector. The device will be operated with a reduced electric field value
in order to simulate very long drift distances of up to about 20 m.
Charge attenuation and amplification will be studied in detail together with the adoption of possible novel technological
solutions. In particular,
several options are being studied for both the HV field shaping electrodes and for the readout devices.

\noindent (4) {\bf The study of LAr TPC prototypes immersed in a magnetic field.}:
Liquid Argon imaging provides very good
tracking with $dE/dx$ measurement, and excellent calorimetric 
performance for contained showers. This allows for a very
precise determination of the energy of the particles in
an event. This is particularly true for electron
showers, which energy is very precisely measured. 

The possibility to complement these features with those
provided by a magnetic field has been considered and would
open new possibilities\cite{Rubbia:2001pk,Rubbia:2004tz}:
(a) charge discrimination,
(b) momentum measurement of particles escaping the detector ($e.g.$ high energy muons),
(c) very precise kinematics, since the measurements are multiple scattering
dominated (e.g. $\Delta p/p\simeq 4\%$ for a track length of $L=12\ m$ and
a field of $B=1T$).

The orientation of the magnetic field is such that the bending direction
is in the direction of the drift where the best spatial resolution is achieved
(e.g. in the ICARUS T600 a point resolution of $400\ \mu m$ was obtained).
The magnetic field is hence perpendicular to the electric field. The Lorentz
angle is expected to be very small in liquid ($e.g.$ $\approx 30 mrad$ at
$E=500\ V/cm$ and $B=0.5T$).
Embedding the volume of Argon into a magnetic field would therefore not
alter the imaging properties of the detector and the measurement of the bending of
charged hadrons or penetrating muons would allow a
precise determination of the momentum and a determination of their charge.

The required magnetic field for charge discrimination for a path $x$ in
the liquid Argon is given by the bending
and the multiple scattering contribution.
The requirement for a $3\sigma$ charge discrimination implies\cite{Rubbia:2004tz}:
\begin{equation}
B\geq \frac{0.2(T)}{\sqrt{x(m)}}
\end{equation}

For long penetrating tracks like muons, a field of $0.1T$ allows
to discriminate the charge for tracks longer than 4 meters. This
corresponds for example to a muon momentum threshold of 800~MeV/c.
Hence, performances are very good, even at very low momenta.

Unlike muons or hadrons, the early showering of electrons 
makes their charge identification difficult. The track length
usable for charge discrimination is limited to a few radiation
lengths after which the showers makes the recognition of
the parent electron more difficult. In practice, charge discrimination
is possible for high fields $x=1X_0 \rightarrow B>0.5T$, $x=2X_0 \rightarrow B>0.4T$,
$x=3X_0 \rightarrow B>0.3T$.
From simulations, we found that the determination
of the charge of electrons of energy in the range between
1 and 5 GeV is feasible with good
purity, provided the field has a strength in the range of 1~T.
Preliminary estimates show that
these electrons exhibit an average curvature 
sufficient to have electron charge discrimination better than
$1\%$ with an efficiency of 20\%. Further studies are on-going.

An R\&D programme to investigate a LAr drift chamber in a magnetic field was started in 2001. 
The goal is to study the drift properties of free electrons in LAr in the presence of a 
magnetic field and to prove that the detection capabilities are not affected. 
We have built a small liquid Argon TPC (width 300~mm, height 150~mm, drift
length 150~mm) and placed
it in the recycled SINDRUM-I magnet\footnote{The magnet was kindly lend to us by PSI, Villigen.}
which allows us to test fields up to 0.5~T. The ongoing test programme includes (1) checking
the basic imaging in B-field (2) measuring traversing and stopping muons (3) test charge
discrimination (4) check Lorentz angle.

\noindent (5) {\bf The further development of the industrial design of a large volume tanker able to operate underground} :
The study initiated with Technodyne UK should be considered as a first ``feasibility'' study, meant to 
select the main issues that will need to be further understood and 
to promptly identify possible ``show-stoppers''.
If successful, we expect to continue this study by more elaborated and detailed industrial design of the
large underground tanker including also the details of the detector instrumentation. 
The cost of the full device will be estimated as well. At this preliminary stage a large mass LAr detector
appears to be a cost effective option.

\noindent (6) {\bf The study of logistics, infrastructure and safety issues related to underground sites} :
We are making preliminary investigations with two ``generic'' geographical configurations: (i) a tunnel-access underground
laboratory, 
(ii) a vertical mine-type-access underground laboratory. Early considerations show that such sites correspond
to interesting complementary options.
Concerning the provision of LAr, a dedicated possibly not underground but nearby air-liquefaction 
plant is foreseen. Technodyne has started 
investigating the technical requirements and feasibility of such a facility.  

These R\&D studies could lead to the necessity of a 10\% full-scale prototype, which could be placed 
at shallow depth, used as an engineering prototype with a physics program on its own\cite{multimw,nufact04}.

\section{Conclusions and outlook}

The Argon Time Projection Chamber (LAr TPC) technology, whose basic R\&D work has been successfully conducted 
by the ICARUS Collaboration, is a mature technique with great potentials. 
We have outlined a strategy for next generation 
experiments on neutrino and astroparticle physics 
based on the use of this technique at different detector mass scales. 

From our general considerations, we can conclude the following points:

(a) a 100 kton liquid Argon TPC based on the tentative design outlined above seems technically
sound and would deliver extraordinary physics output (sometimes called ``megaton physics"). 
This device could effectively compete with giant 0.5-1 Megaton water Cerenkov detectors being proposed for 
future precision studies of the neutrino mixing matrix and for nucleon decay searches\cite{Rubbia:2004yq}.
Coupled to future Superbeams\cite{Ferrari:2002yj}, Betabeams or Neutrino Factories\cite{Rubbia:2001pk,Bueno:2001jd}
it could greatly improve our understanding of the mixing matrix in the lepton sector with
the goal of measuring the CP-phase, and in parallel it would
allow to conduct astroparticle experiments of unprecedented sensitivity\cite{multimw,Gil-Botella:2004bv}.
The main design features include the possibility of
a bi--phase operation with charge amplification for long drift distances,
an imaging plus scintillation plus (possibly) Cerenkov readout for improved physics performance, and
a very large boiling industrial cryostat (LNG technology).
The main issues are related to finding a practical underground location --
deep or shallow depth depending of physics goals -- and an appropriate funding on the scale. 
A rough estimate of the cost has been elaborated\cite{nufact04}. 

(b) a 10\% full-scale, cost effective prototype of the design outlined above on the scale of
10 kton could be readily envisaged as an engineering design test with a physics program
of its own, directly comparable to that of Superkamiokande. This would provide a direct
and probably final demonstration of the advantages of a very large scale liquid Argon TPC
compared to other existing or planned techniques. A rough estimate of the cost has been 
elaborated\cite{nufact04}. 

(c) A 100 ton detector in a near-site of a long-baseline facility is a straight forward and
very desirable application of the technique. For example, the approved T2K experiment in Japan
or other beams in the US might provide the ideal conditions to study with high statistical
accuracy neutrino interactions on liquid Argon in the very important energy range around 1 GeV.
This is a mandatory step in order to be able to handle high statistics provided by large detectors.
A prototype of a 100 ton detector could provide a tool to study calorimetric (electromagnetic
and hadronic) response in a charged particle beam or be readily placed in an existing
neutrino beam.

Work is in progress along these lines of thoughts. 

There is a high degree of interplay and a strong
synergy between small and large mass scale apparatuses, the very large
detector needing the small one in order to best exploit the measurements with high statistical precision
that will be possible with a large mass. We believe (and hope) 
that small and very large LAr detectors will play significant and important roles in the future.

\section{Acknowledgments}
We thank I. Gil-Botella for a careful reading of this manuscript.
Part of this work was supported by ETH/Z\"urich and Swiss National Science Foundation.


\begin{thebibliography}{99}
                 
\bibitem{intro1} C. Rubbia, 
 CERN--EP/77--08, (1977).

\bibitem{t600paper}
S.~Amerio {\it et al.} [ICARUS Collaboration],
Nucl. Inst. Meth., A527 (2004) 329-410  and
references therein.

\bibitem{3tons}
P.~Benetti {\it et al.} [ICARUS Collaboration],
Nucl.\ Instrum.\ Meth.\ A { 332}, (1993) 395. 

\bibitem{Cennini:ha}
P.~Cennini {\it et al.}, [ICARUS Collaboration],
Nucl.\ Instrum.\ Meth.\ A { 345}, (1994) 230.

\bibitem{50lt}
F.~Arneodo {\it et al.}  [ICARUS Collaboration],
arXiv:hep-ex/9812006.

\bibitem{Amoruso:2003sw}
S.~Amoruso {\it et al.}  [ICARUS Collaboration],
arXiv:hep-ex/0311040. The European Physical Journal C,  Eur. Phys. J. C 33, 233-241 (2004).

\bibitem{Amoruso:2004dy}
S.~Amoruso {\it et al.}  [ICARUS Collaboration],
Nucl.\ Instrum.\ Meth.\ A {\bf 523} (2004) 275.

\bibitem{gg2}
M.~Antonello {\it et al.}  [ICARUS Collaboration],                             
                         Nucl.\ Instrum.\ Meth.\ A516 (2004) 348-363.                         
                         
\bibitem{gg3}
S.~ Amoruso {\it et al.}  [ICARUS Collaboration],
                         Nucl.\ Instrum.\ Meth.\ A516 (2004) 68-79.
                         
\bibitem{gg1}
F.~Arneodo {\it et al.}  [ICARUS Collaboration],
Nucl.\ Instrum.\ Meth.\ A 508, 287 (2003).

\bibitem{Itow:2001ee}
Y.~Itow {\it et al.},
arXiv:hep-ex/0106019.

\bibitem{Ahn:2002up}
M.~H.~Ahn {\it et al.}  [K2K Collaboration],
Phys.\ Rev.\ Lett.\  {\bf 90} (2003) 041801
[arXiv:hep-ex/0212007].

\bibitem{Rubbia:2001pk}
A.~Rubbia,
arXiv:hep-ph/0106088.

\bibitem{Rubbia:2004yq}
A.~Rubbia,
arXiv:hep-ph/0407297.

\bibitem{Rubbia:2004tz}
A.~Rubbia,
arXiv:hep-ph/0402110.

\bibitem{Technodyne} Technodyne International Limited,
Unit 16 Shakespeare Business Center Hathaway Close,
Eastleigh, Hampshire, SO50 4SR, see {\it{http://www.technodyne.co.uk}}

\bibitem{Sauli:qp}
F.~Sauli,
Nucl.\ Instrum.\ Meth.\ A { 386} (1997) 531.

\bibitem{Jeanneret:mr}
P.~Jeanneret, J.~Busto, J.~L.~Vuilleumier, A.~Geiser, V.~Zacek, H.~Keppner and R.~de Oliveira,
Nucl.\ Instrum.\ Meth.\ A { 500} (2003) 133.

\bibitem{dmrd} R.~Chandrasekharan, M.~Messina, P.~Otiougova, P.~Picchi, F.~Pietropaolo, A.~Rubbia,
in preparation.

\bibitem{multimw}
A.~Ereditato and A.~Rubbia,
``Ideas for a next generation liquid Argon TPC detector for neutrino physics and nucleon decay searches,''
Memorandum submitted to the CERN SPSC Villars session, April 2004. 

\bibitem{Ferrari:2002yj}
A.~Ferrari, A.~Rubbia, C.~Rubbia and P.~Sala,
New J.\ Phys.\   4 (2002) 88
[arXiv:hep-ph/0208047].

\bibitem{Bueno:2001jd}
A.~Bueno, M.~Campanelli, S.~Navas-Concha and A.~Rubbia,
Nucl.\ Phys.\ B { 631} (2002) 239
[arXiv:hep-ph/0112297]; Nucl.\ Phys.\ B { 589} (2000) 577
[arXiv:hep-ph/0005007].

\bibitem{Gil-Botella:2004bv}
I.~Gil-Botella and A.~Rubbia,
JCAP {\bf 0408} (2004) 001
[arXiv:hep-ph/0404151]; JCAP {\bf 0310} (2003) 009
[arXiv:hep-ph/0307244].

\bibitem{nufact04}
A.~Ereditato and A.~Rubbia,
``Liquid Argon TPC: mid \& long term strategy and on-going R\&D'', To appear
in {\it  International Conference on NF and Superbeam, NUFACT04, Osaka, Japan, July 2004}.

\end{thebibliography}
\end{document}